\documentstyle[amssymbols]{article}

\author{Aleksandar Ignjatovic and Arun Sharma}
\title{Some Applications of Logic to Feasibility in Higher Types}
\begin{document}

\newcommand{\arl}[2]{$\blacktriangleleft\!\!\!\frac{\ \ \mbox{ #1}\ \ }{\ \ 
\mbox{ #2}\ \ }$}
\newcommand{\arr}[2]{$\frac{\ \ \mbox{ #1}\ \ }{\ \ 
\mbox{ #2}\ \ }\!\!\!\blacktriangleright$}
\newcommand{\llangle}{\langle\kern-.25em\langle}
\newcommand{\rrangle}{\rangle\kern-.25em\rangle}
\newcommand{\LL}{L^2_b}
\newcommand{\es}{\emptyset}
\newcommand{\ek}{\approx}
\newcommand{\mem}{\:\varepsilon\:}
\newcommand{\bbv}{\bigvee\negthinspace\negthinspace
\negthinspace\negthinspace\bigvee}
\newcommand{\bbw}{\bigwedge\negthinspace\negthinspace
\negthinspace\negthinspace\bigwedge}
\newcommand{\ekh}{\hat{\approx}}
\newcommand{\memh}{\ \hat{\varepsilon}\ }
\newcommand{\notm}{\not\negthinspace\varepsilon\ }
\newcommand{\si}[1]{\{ #1 \} }
\newcommand{\pr}[2]{\{ #1 , #2 \} }
\newcommand{\epsi}[3]{E( #1 \cap #2 ,#3 )}
\newcommand{\tim}[3]{(#1 \cap #2 ) \times #3 }
\newcommand{\val}[1]{{\mathcal{V}}[ #1 ] }
\newcommand{\valt}[1]{{\mathcal{V}}[ #1 ]=\top}
\newcommand{\valf}[1]{{\mathcal{V}}[ #1 ]=\bot}
\newcommand{\un}[1]{\underline{#1 }}
\newtheorem{theorem}{Theorem}
\newtheorem{lemma}[theorem]{Lemma}
\newtheorem{definition}{Definition}
\newtheorem{conjecture}[theorem]{Conjecture}%
\newtheorem{corollary}[theorem]{Corollary}%
\newtheorem{proposition}[theorem]{Proposition}%
\newtheorem{question}[theorem]{Question}%
\newcommand{\dotminus}{\mathbin{\mathchoice%
{\buildrel .\lower.6ex\hbox{\vphantom{.}} \over {\smash-}}%
{\buildrel .\lower.6ex\hbox{\vphantom{.}} \over {\smash-}}%
{\buildrel .\lower.4ex\hbox{\vphantom{.}} \over {\smash-}}%
{\buildrel .\lower.3ex\hbox{\vphantom{.}} \over {\smash-}}}}%

\newcommand{\halfof}[1]{\lfloor {\textstyle\frac{1}{2}} #1 \rfloor}

\newcommand{\bbox}
	{\hbox{
	    \vbox{\hrule height.2ex
		  \hbox{\vrule height1.5ex width0.2em
			\kern .32em
			\vrule height1.5ex width0.2em}
		  \hrule height.2ex}} \kern .1em}
\newcommand{\pbox}
	{\hbox{
	    \vbox{\hrule height.03ex
		  \hbox{\vrule height1.5ex width0.05em
			\kern .32em
			\vrule height1.5ex width0.05em}
		  \hrule height.03ex}} \kern .1em}

%
%
\def\LFF{\mbox{${\rm L}_{\rm FF}$}}
\def\boxit#1{\vbox{\hrule\hbox{\vrule\kern3pt\vbox{\kern3pt#1\kern3pt}\kern3pt\vrule}\hrule}}
\def\BFF{{\bf BFF}}
\def\PV{\mbox{${\rm PV}_1$}}
\def\S{\mbox{${\rm S}^1_2$}}
\def\PVF{\mbox{${\rm PVF}_2$}}%
\def\BFFM{\mbox{${\rm BFF}^-$}}%
\def\BR{\mbox{$({\bf \Sigma}^b_1-LPIND)$}}
\def\BS{\mbox{$({\bf \Sigma}^b_1-PIND)$}}
\def\BSi{\mbox{$({\bf \Sigma}^b_i-PIND)$}}
\def\BSH{\mbox{$({\bf \Sigma}^b_1({\cal{B}})-PIND)$}}
\def\BPT{\mbox{${\cal B}_{\rm  P}$}}
\def\Ap{\mbox{${\it Ap}$}}
\def\SqBd{\mbox{${\it SqBd}$}}
\def\SBS{\mbox{$({\it s}-{\bf \Sigma}^b_1({\cal{B}})-PIND)$}}
\def\nBS{\mbox{$({\it n}-{\bf \Sigma}^b_1({\cal{B}})-PIND)$}}
\def\prefixF{\mbox{${(\exists\vec{\bf{z}}^F \le \vec{\bf{t}}^F)}$}}
\def\prefixK{\mbox{${(\exists\vec{\bf{z}}^K \le \vec{\bf{t}}^K)}$}}
\def\prefixH{\mbox{${(\exists\vec{\bf{z}}^H \le \vec{\bf{t}}^H)}$}}
\def\prefixg{\mbox{${(\exists\vec{\bf{z}}^G \le \vec{\bf{t}}^G)}$}}
\def\prefixGj{\mbox{${(\exists\vec{\bf{z}}^{G_j} \le \vec{\bf{t}}^{G_j})}$}}
\def\prefixG1{\mbox{${(\exists\vec{\bf{z}}^{G_1} \le \vec{\bf{t}}^{G_1})}$}}
\def\prefixGl{\mbox{${(\exists\vec{\bf{z}}^{G_l} \le \vec{\bf{t}}^{G_l})}$}}

\def\proves{\vdash}
\newcommand{\BASIC}{BASIC}
\newcommand{\SS}{{\bf S}^1_2}
\newcommand{\RR}{{\bf R}^1_2}
\newcommand{\RRi}{{\bf R}^i_2}
\newcommand{\SSi}{{\bf S}^i_2}
\newcommand{\PIND}{\mbox{-PIND}}
\newcommand{\LIND}{\mbox{-LIND}}
\newcommand{\LPIND}{\mbox{-LPIND}}
\newcommand{\IND}{\mbox{-IND}}
\newcommand{\LENP}{\mbox{LenP}}
\newcommand{\LENMINUS}{\mbox{LenMinus}}
\newcommand{\SPtwo}{\mbox{SubPower2}}
\newcommand{\SE}{\mbox{SubExp}}
\newcommand{\LSP}{\mbox{LSP}}
\newcommand{\MSP}{\mbox{MSP}}
\newcommand{\impl}{\rightarrow}
\newcommand{\tfi}{Thm_T(\lceil \varphi \rceil)}
\newcommand{\lrc}{\begin{eqnarray*}
&F(\vcc,0)  =  G(\vcc)&\\
&F(\vcc,u)  =  H(\vcc,u,F(\vcc,\half{u})),&\;\;u >0,\\
&|F(\vcc,u)|  \leq  |K(\vcc,u)|&
\end{eqnarray*}}
\newcommand{\bigor}{\bigvee}                    
\newcommand{\bigand}{\bigwedge}                 
\newcommand{\pf}{p(\vec{a},b_1,\dots,b_i)}
\newcommand{\pfa}{p(\vec{a})}
\newcommand{\pfb}{p(b_0,\dots,b_k)}
\newcommand{\pfab}{p(\vec{a},b_0,\dots,b_k)}
\newcommand{\ov}{\overline}
\newcommand{\co}[1]{\lgn #1 \rgn}
\newcommand{\PF}{\noindent {\bf Proof:~~~}}
\newcommand{\two}[1]{\underline{2^{#1}}}
\newcommand{\dva}[1]{2^{#1}_{|t|}}
\newcommand{\exy}[2]{2^{#1}_{| #2 |}}
\newcommand{\dub}[1]{ 2 \cdot #1}
\newcommand{\num}[1]{\underline{#1}}
\newcommand{\liff}{\leftrightarrow}
\newcommand{\ar}{\Longrightarrow}
\newcommand{\li}{\ \, \land\, \ }
\newcommand{\lili}{\ \, \lor \, \ }
\newcommand{\res}{\negthinspace\restriction\negthinspace}
\newcommand{\paref}[1]{{\rm(\ref{#1})}}
\newcommand{\qed}{\hfill$\blacksquare$\kern2pt}
\newcommand{\pair}[2]{\llangle #1,#2 \rrangle}
\newcommand{\vccc}{\vec{f},\vec{x}}
\newcommand{\vcc}{f,\vec{x}}
\newcommand{\avcc}{|\vec{f}|,|\vec{x}|}
\newcommand{\avc}{|f|,|\vec{x}|}
\newcommand{\PP}{P(\avcc)}
\newcommand{\Pp}{P(\avc)}
\newcommand{\FF}{F(\vccc)}
\newcommand{\Ff}{F(\vcc)}
\newcommand{\BFS}{{\bf S}^1_2}
\newcommand{\BFR}{{\bf R}^1_2}
\newcommand{\nf}{|f|(|x|)}
\newcommand{\half}[1]{\lfloor {\textstyle \frac{1}{2}} #1 \rfloor}
\newcommand{\fr}[2]{\lfloor \frac{\textstyle #1}{\textstyle #2} \rfloor}
\newcommand{\tr}[2]{{\cal #1}_{#2}(f,\vec{x})}
\newcommand{\trg}[3]{{\cal #1}_{#2}(f,\vec{x} #3)}
\newcommand{\trs}[2]{\overline{{\cal #1}}_{G,H}(f,\vec{x}, #2)}
\newcommand{\ru}{\frac{\textstyle \Gamma, \neg (b_i \leq t(\vec{b})), 
A(b_i)}{\textstyle \Gamma, (\forall x \leq t(\vec{b}))A(x)}}
\newcommand{\ruu}{\frac{\textstyle \Gamma, A(b_i)}{\textstyle \Gamma, 
(\forall x) A(x)}}
\newcommand{\rU}{\frac{\textstyle \Gamma, A(f_i)}{\textstyle \Gamma, 
(\forall f^*_i) A(f^*_i)}}
\newcommand{\re}{\frac{\textstyle \Gamma, s\leq t \& A(s)}{\textstyle 
\Gamma, (\exists x\leq t)A(x)}}
\newcommand{\reu}{\frac{\textstyle \Gamma, A(s)}{\textstyle \Gamma, 
(\exists x) A(x)}}
\newcommand{\rE}{\frac{\textstyle \Gamma, A(t)}{\textstyle \Gamma, 
(\exists f^*_i) A(f^*_i)}}
\newcommand{\lind}[3]{\frac{\textstyle \Gamma(\vec{\alpha}), #1 \; #2 
(0,\vec{\alpha}) \;\;\;\;\;\;\;\; \Gamma(\vec{\alpha}), #1 \; \neg #2 
(b_i,\vec{\alpha}),\; #2 (b_i+1,\vec{\alpha})}{\textstyle 
\Gamma(\vec{\alpha}), #1 \; #2 (| #3 (\vec{\alpha})|)}}
\newcommand{\A}{{\cal A}}
\newcommand{\maxindex}{\mbox{\sc maxindex}}
\newcommand{\blocksize}{\mbox{\sc blocksize}}
\newcommand{\vecc}{\overline}
\newcommand{\ocram}{\mbox{\sc ocram}}
\newcommand{\cl}{\noindent {\bf Claim.  }}
\newcommand{\restrict}{\ |\grave{}\ }
\newcommand{\xvec}{x_1,\ldots,x_n}
\newcommand{\Q}{{\cal Q}} 
\def\bw{\bigwedge\kern-.650em \bigwedge}
\def\slv{/\kern-.670em \vdash}
\def\Nu{\mbox{\large $\nu$}}
\newcommand{\BIT}{\mbox{BIT}}
\newcommand{\RBIT}{\mbox{RBIT}}
\def\monus{\mathbin{\mathchoice%
  {\vbox{\rlap{\hskip 2.55pt\raise 3.5pt\hbox{.}}}\mathord-}%
  {\vbox{\rlap{\hskip 2.55pt\raise 3.5pt\hbox{.}}}\mathord-}%
  {\vbox{\rlap{\hskip 1.5pt\raise 2.6pt\hbox{.}}}\mathord-}%
  {\vbox{\rlap{\hskip 1pt\raise 1.9pt\hbox{.}}}\mathord-}%
}}

\include{mssymb}
\include{amssymbols.sty}
%
%
\catcode`\@=11

\def\newdefinition#1{\@ifnextchar[{\@odfn{#1}}{\@ndfn{#1}}}

\def\@ndfn#1#2{\@ifnextchar[{\@xndfn{#1}{#2}}{\@yndfn{#1}{#2}}}

\def\@xndfn#1#2[#3]{\expandafter\@ifdefinable\csname #1\endcsname
{\@definecounter{#1}\@addtoreset{#1}{#3}%
\expandafter\xdef\csname the#1\endcsname{\expandafter\noexpand
  \csname the#3\endcsname \@dfncountersep \@dfncounter{#1}}%
\global\@namedef{#1}{\@dfn{#1}{#2}}\global\@namedef{end#1}{\@enddefinition}}}

\def\@yndfn#1#2{\expandafter\@ifdefinable\csname #1\endcsname
{\@definecounter{#1}%
\expandafter\xdef\csname the#1\endcsname{\@dfncounter{#1}}%
\global\@namedef{#1}{\@dfn{#1}{#2}}\global\@namedef{end#1}{\@enddefinition}}}

\def\@odfn#1[#2]#3{\expandafter\@ifdefinable\csname #1\endcsname
  {\global\@namedef{the#1}{\@nameuse{the#2}}%
\global\@namedef{#1}{\@dfn{#2}{#3}}%
\global\@namedef{end#1}{\@enddefinition}}}

\def\@dfn#1#2{\refstepcounter
    {#1}\@ifnextchar[{\@ydfn{#1}{#2}}{\@xdfn{#1}{#2}}}

\def\@xdfn#1#2{\@begindefinition{#2}{\csname the#1\endcsname}\ignorespaces}
\def\@ydfn#1#2[#3]{\@opargbegindefinition{#2}{\csname
       the#1\endcsname}{#3}\ignorespaces}

\def\@dfncounter#1{\noexpand\arabic{#1}}
\def\@dfncountersep{.}
\def\@begindefinition#1#2{\rm \trivlist \item[\hskip \labelsep{\bf
\hskip\parindent #2.\ #1.}]} \def\@opargbegindefinition#1#2#3{\rm
\trivlist
      \item[\hskip \labelsep{\bf \hskip\parindent #2.\ #1\ \rm
(#3).}]} \def\@enddefinition{\endtrivlist}

\def\newproof#1{\@ifnextchar[{\@opf{#1}}{\@npf{#1}}}

\def\@npf#1#2{\@ifnextchar[{\@xnpf{#1}{#2}}{\@ynpf{#1}{#2}}}

\def\@xnpf#1#2[#3]{\expandafter\@ifdefinable\csname #1\endcsname
{\@definecounter{#1}\@addtoreset{#1}{#3}%
\expandafter\xdef\csname the#1\endcsname{\expandafter\noexpand
  \csname the#3\endcsname \@pfcountersep \@pfcounter{#1}}%
\global\@namedef{#1}{\@pf{#1}{#2}}\global\@namedef{end#1}{\@endproof}}}

\def\@ynpf#1#2{\expandafter\@ifdefinable\csname #1\endcsname
{\@definecounter{#1}%
\expandafter\xdef\csname the#1\endcsname{\@pfcounter{#1}}%
\global\@namedef{#1}{\@pf{#1}{#2}}\global\@namedef{end#1}{\@endproof}}}

\def\@opf#1[#2]#3{\expandafter\@ifdefinable\csname #1\endcsname
  {\global\@namedef{the#1}{\@nameuse{the#2}}%
\global\@namedef{#1}{\@pf{#2}{#3}}%
\global\@namedef{end#1}{\@endproof}}}

\def\@pf#1#2{\refstepcounter
    {#1}\@ifnextchar[{\@ypf{#1}{#2}}{\@xpf{#1}{#2}}}

\def\@xpf#1#2{\@beginproof{#2}{\csname the#1\endcsname}\ignorespaces}
\def\@ypf#1#2[#3]{\@opargbeginproof{#2}{\csname
       the#1\endcsname}{#3}\ignorespaces}

\def\@pfcounter#1{\noexpand\arabic{#1}}
\def\@pfcountersep{.}
\def\@beginproof#1#2{\rm \trivlist \item[\hskip \labelsep{\bf
\hskip\parindent #1.}]} \def\@opargbeginproof#1#2#3{\rm \trivlist
      \item[\hskip \labelsep{\bf \hskip\parindent #1 of #3.}]}
\def\@endproof{\hspace*{\fill}{\vrule height4pt width4pt depth0pt}\endtrivlist}

\newproof{proof}{Proof}
\newproof{proofofl}{Proof of Lemma}
\newproof{sketchofproof}{Sketch of Proof}

%
%

\newcommand{\pind}[3]{\frac{\textstyle \Gamma(\vec{\alpha}), #1 \; #2 (0,\vec{\alpha}) \;\;\;\;\;\;\;\; \Gamma(\vec{\alpha}), 
#1 \; \neg #2 (\half{b_i},\vec{\alpha}),
\; #2 (b_i,\vec{\alpha})}{\textstyle \Gamma(\vec{\alpha}), #1 \; #2 (#3 (\vec{\alpha}),\vec{\alpha})}}

\def\BFF{\mbox{${\cal B}$}}
\def\LFF{\mbox{${\rm L}_{\cal B}$}}
\def\PV{\mbox{${\rm PV}_1$}}
\def\S{\mbox{${\rm S}^1_2$}}
\def\PVF{\mbox{$({\rm QF}({\cal{B}})-PIND)$}}
\def\BFFM{\mbox{${\cal B}$}}
\def\BS{\mbox{$({\bf \Sigma}^b_1-PIND)$}}
\def\BSi{\mbox{$({\bf \Sigma}^b_i-PIND)$}}
\def\BSH{\mbox{$({\bf \Sigma}^b_1({\cal{B}})-PIND)$}}
\def\BPT{\mbox{${\cal B}_{\rm  P}$}}
\def\Ap{\mbox{${\it Ap}$}}
\def\SqBd{\mbox{${\it SqBd}$}}
\def\SBS{\mbox{$({\it s}-{\bf \Sigma}^b_1({\cal{B}})-PIND)$}}
\def\nBS{\mbox{$({\it n}-{\bf \Sigma}^b_1({\cal{B}})-PIND)$}}
\def\prefixF{\mbox{${(\exists\vec{\bf{z}}^F \le \vec{\bf{t}}^F)}$}}
\def\prefixK{\mbox{${(\exists\vec{\bf{z}}^K \le \vec{\bf{t}}^K)}$}}
\def\prefixH{\mbox{${(\exists\vec{\bf{z}}^H \le \vec{\bf{t}}^H)}$}}
\def\prefixg{\mbox{${(\exists\vec{\bf{z}}^G \le \vec{\bf{t}}^G)}$}}
\def\prefixGj{\mbox{${(\exists\vec{\bf{z}}^{G_j} \le \vec{\bf{t}}^{G_j})}$}}
\def\prefixG1{\mbox{${(\exists\vec{\bf{z}}^{G_1} \le \vec{\bf{t}}^{G_1})}$}}
\def\prefixGl{\mbox{${(\exists\vec{\bf{z}}^{G_l} \le \vec{\bf{t}}^{G_l})}$}}
\def\proves{\vdash}

\newtheorem{remark}{Remark}
\maketitle
\begin{abstract}
While it is commonly accepted that computability on a Turing 
machine in polynomial time represents a correct formalisation of the 
notion of a {\em feasibly computable} function, 
there is no similar agreement on how to extend this notion on 
{\em functionals}, i.e., what functionals should be 
considered feasible. One possible paradigm was introduced by 
K.~Mehlhorn in \cite{Mehl}, 
who extended Cobham's definition of feasible functions to type 2 
functionals. Subsequently, this class of functionals (with inessential 
changes of the definition) was studied by Townsand (\cite{Town}) who 
calls this class {\bf POLY}, and by 
Cook and Kapron (\cite{KapCook}) who call the same class 
{\em basic feasible functionals}. Cook and Kapron gave an oracle Turing 
machine model characterization of this class.
In this paper we demonstrate that the class of basic feasible 
functionals has recursion theoretic 
properties which naturally generalize the corresponding properties 
of the class of feasible functions, thus giving furter evidence that the notion of 
feasibility of functionals mentioned above is correctly chosen. We
also improve the Kapron-Cook result on machine representation.

Our proofs are based on essential applications of logic. We introduce 
a weak fragment of second order arithmetic with second order variables 
ranging over functions from ${\Bbb N}^{\Bbb N}$ which suitably characterizes 
basic feasible functionals, and show that it is 
a useful tool for investigating the properties of basic feasible 
functionals. 
In particular we provide an example how one can extract feasible ``programs'' 
from mathematical proofs which use non-feasible functionals. 
\end{abstract}
\section{Introduction}

\subsection{Recursion Theory of Feasible Functionals}

In \cite{Cobh}  Cobham has established the following fundamental result: functions 
computable on a Turing machine in polynomial time are exactly
functions which can be obtained from the basic functions\footnote{Recall 
that $|x|$ denotes the length of the
binary representation of $x$, that is to say $\lceil (log_2 x) + 1\rceil$,
with $|0|=0$;  $\half {x}$ denotes the greatest integer less than or
equal to $x/2$; $x\# y$ is equal to $2^{|x|\cdot |y|}$.} 
${\bf o}(x)=0, s_0(x)=2x, s_1(x)=2x+1, i^n_k(x_1,\ldots x_n)=x_k$ and
$x\# y$ by composition and by limited recursion on notation: 
\begin{eqnarray}
f(\vec{x},0) & = & g(\vec{x})\label{i}\\
f(\vec{x},2y) & = & h_1(\vec{x},y,f(\vec{x},y)),\label{ii}\ \ \ \ \ y >0\\
f(\vec{x},2y+1) & = & h_2(\vec{x},y,f(\vec{x},y))\label{iii}\\
|f(\vec{x},y)| & \leq & |k(\vec{x},y)|\label{iv}
\end{eqnarray}
It is easy to see that the same class of functions is obtained if 
we replace the
schema of limited recursion on notation by a schema which instead of 
\paref{ii} and \paref{iii} has
\begin{equation}\label{v}
f(\vec{x},y) =  h(\vec{x},y,f(\vec{x},\lfloor 
\mbox{$\frac{y}{2}$} \rfloor)),\ \ \ \ \ y >0\\
\end{equation}
Also, instead of condition \paref{iv}, we can  take the 
condition
\begin{equation}
|f(\vec{x},y)|  \leq q(|\vec{x}|,|y|)
	\label{vi}
\end{equation}
where $q$ is a polynomial with  natural coefficients.

Finally, the same set can be obtained if the schema of limited recursion 
on notation is replaced by the following version of the  schema of 
primitive recursion, in which $p$ and $q$ are polynomials 
and $f^*$ is an auxiliary primitive recursive function.
\begin{eqnarray}
&f^*(\vec{y},0)  =  g(\vec{y})&
	\label{la} \\
&f^*(\vec{y},x+1)  =  h(\vec{y},x,f^*(\vec{y},x))&
	\label{lalaa} \\
&(\forall x\leq p(|\vec{y}|)) (|f^*(\vec{y},x)|  \leq  q(|\vec{y}|))&
	\label{lalala} \\
&f(\vec{y})  =  f^*(\vec{y},p(|\vec{y}|))&
	\label{lalalala}
\end{eqnarray}

From the foundational point of view we note that   
conditions \paref{vi} and \paref{lalala}   limit the growth rate of the function 
being defined using functions which are particularly simple 
to compute (polynomials), in the sense that
their definition  does not involve any recursion.

\begin{definition} {\sl 
A function $F: ({\Bbb N}^{\Bbb N})^k \times {\Bbb N}^l \rightarrow {\Bbb N}$
is called {\it a type 2 functional}  of {\it rank} $(k,l)$.}
\end{definition}

One of our aims is to find analogues of the above schemata for a class of
functionals which can be seen as natural extension of the class of 
feasible functions.

\begin{definition} {\sl 
A functional of rank $(k,l)$ is obtained by
\begin{description}
\item{\it Functional composition} from $H(\vec{f},\vec{y},\vec{x}),\ 
G_1(\vccc), \ldots, G_l(\vccc)$ if
\[
\FF = H(\vec{f}, G_1(\vccc), \ldots, 
G_l(\vccc),\vec{x})
\]
\item{\it Expansion} from $G(\vccc)$ if
\[
F(\vec{f},\vec{g},\vec{x},\vec{y}) = G(\vccc)
\]
\item{\it Functional substitution} from 
$H(\vec{g},\vec{f},\vec{x}),G_1(\vccc,y), \ldots, G_l(\vccc,y)$ if
\[
\FF = H(\lambda y.G_1(\vccc,y), \ldots, 
\lambda y.G_l(\vccc,y),\vccc)
\]
\item{\it Limited recursion on notation} from $G,H_1,H_2,K$ if the following
hold for all $\vec f \in ({\Bbb N}^{\Bbb N})^k$ and all $\vec x \in {\Bbb 
N}^l$:
\begin{eqnarray}
F(\vccc,0) & = & G(\vccc)\label{z}\\
F(\vccc,2y) & = & H_1(\vccc,y,F(\vccc,y)),\ \ \ \ \ y >0\label{zz}\\
F(\vccc,2y+1) & = & H_2(\vccc,y,F(\vccc,y)),\label{zzz}\\
|F(\vccc,y)| & \leq & |K(\vccc,y)|\label{zzzz}
\end{eqnarray}
\end{description}
}\end{definition}
The last condition is equivalent to the condition 
$F(\vccc,y)  \leq K^\ast(\vccc,y)$ 
for $K^\ast = (1 \# K(\vccc,y))\dotminus 1$.

Townsend  (\cite{Town}) considered the least class of functionals which 
contains the polynomial time computable functions, the application functional 
 $\Ap$ defined by $\Ap(f,x) = f(x)$, which is closed under expansion,
functional composition, functional substitution and limited recursion on 
notation. He has also shown that the scheme of functional
substitution is redundant; thus we introduce the following
appropriately modified definition.
\begin{definition} {\sl 
The class $\BFF$ of {\it basic feasible functionals} is the least
class of functionals which contains initial functions 
${\bf o}(x)=0$, $s_0(x)=2x$, $s_1(x)=2x+1$, $i^n_k(x_1,\ldots x_n)=x_k$, 
$x\# y$ and the application functional $Ap(f,x)=f(x)$, 
which is closed under expansion,
functional composition and limited recursion on notation.
}\end{definition}

The class of basic feasible functionals extends the class of 
feasible functions in a ``minimal way''; it is obtained using 
essentially  the same closure conditions as the class of feasible 
functions (taking into account 
that the schema of functional substitution is redundant and so it can be 
omitted). The only difference is addition of the application functional 
to the set of basic functions. However, in any reasonable model of 
feasibility in higher types such functional must be considered feasible.
Also, it is easy to see 
that the only functions which belong to the class of the basic feasible 
functionals are in fact just feasible functions.

As before, recursion on notation of type \paref{zz} together 
with \paref{zzz} can be replaced with a single condition
$$F(\vccc,y) =  H(\vccc,y,F(\vccc,\half{y})).$$

However, the growth rate of a basic feasible functional clearly cannot be 
majorized by a first order polynomial. Thus, we cannot expect to have 
simple analogues for schemas involving polynomial bounds or limited 
primitive recursion. The same applies to the machine models of 
feasibility in higher types. This is why we need the following 
definitions introduced by Kapron in \cite{KapCook}.

\begin{definition} {\sl 
The  functional of type ${\Bbb N}^{\Bbb N}\rightarrow {\Bbb N}^{\Bbb N}$ 
defined such that $f \mapsto |f|$ where
\[
|f|(x) = \max_{|y| \le x} |f(y)|
\]
is called {\it the norm functional}; the function $|f|$ is called 
the {\it norm} of the function $f$.
}\end{definition}

Kapron has shown that the functional of type (1,1) such that
$\langle f,x \rangle
\mapsto |f|(|x|)$ is not basic feasible and that, in fact, it cannot be
majorized by any basic feasible functional. 

The class of {\it second-order polynomials} was also introduced in
\cite{KapCook}. Second order polynomials play quite the same role which
first order polynomials play for feasible functions of type 
${\Bbb N}^k \rightarrow {\Bbb N}$. However, there is a major difference
in their nature which greatly complicates their applications: while
first order polynomials are themselves feasible functions, the above
result of Kapron implies that the second order polynomials are {\it not}
feasible functionals.

\begin{definition} {\sl 
Let $x_0, x_1,\ldots$ and $f_0, f_1,\ldots$ be sets of first and second order 
variables respectively; then  the set of {\it second order polynomials} in 
$|f_0|, |f_1|,\ldots$ and $|x_0|, |x_1|,\ldots$ is defined 
inductively as the least set of terms of the language $L_P$ containing 
constants $\underline{n}$ for each natural number $n$ 
and all terms $|x_0|, |x_1|,\ldots$ and which satisfies the following closure 
condition:
if $P,Q$ are second-order polynomials and $f_i$ is a second-order
variable, then $P+Q$, $P \cdot Q$ and $|f_i|(P)$ are also second-order 
polynomials.}
\end{definition}
Clearly, second order polynomials take into account the growth rate of 
functions. However, 
they cannot be considered to be feasible: to compute $|f|(|x|)$, we must 
compute $|f(y)|$ for all $y$ such that $|y|\leq |x|$, which is 
exponentially many values of $y$ (in $|x|$). Kapron's  
result showing  that the norm is not basic feasible implies that the same 
holds of second order polynomials.

We can now state the main results of this paper, which are the best 
possible analogues of the corresponding first order theorems. Due to the 
mentioned difficulty (non-feasibility of the second order polynomials), 
the proofs of these analogues use formal logic as an essential tool.
\begin{definition}\label{pbr} {\sl 
Let $Q(\avc)$ be a
second order polynomial and let 
$G(\vcc)$ and  $H(\vcc,z,y)$
be two basic feasible functionals. Assume that the functional   
$F(\vcc,y)$ satisfies 
\begin{eqnarray}
&F(\vcc,0) = G(\vcc)&\nonumber\\
&F(\vcc,y) = H(\vcc,F(\vcc,\half{y}),y),&\nonumber\\
&|F(\vcc, y)| \leq Q(|f|, |\vec{x}|,|y|))&\label{kkk}
\end{eqnarray}
Then we say that $F$ is defined from functionals $G,H$ 
by {\it polynomially bounded recursion on notation with the bound} $Q$.
}\end{definition}

\noindent {\bf Theorem}:~~{\sl
Assume that the functional $F(\vcc,y)$ is defined from the functionals 
$G(\vcc)$ and $H(\vcc,z,y)$ by polynomially bounded
recursion on notation with the bound $Q(|f|, |\vec{x}|,|y|)$. Then
$F(\vcc,y)$ is a basic feasible functional.}

\begin{definition}\label{pbpl} {\sl 
Let $P(|f|, |\vec{x}|)$ and $Q(\avc)$ be two
second order polynomials and let 
$G(\vcc)$ and  $H(\vcc,y,z)$
be two basic feasible functionals. Assume that the functional   
$F^\ast(\vcc,y)$ satisfies 
\begin{eqnarray*}
&F^\ast(\vcc,0) = G(\vcc);&\\
&F^\ast(\vcc,y+1) = H(\vcc,F^\ast(\vcc,y),y);&\\
&(\forall y \leq P(|f|, |\vec{x}|))(|F^\ast (\vcc, y)|\leq Q(|f|, 
|\vec{x}|));&\label{ocond}\\
&(\forall y) (y \geq P(\avc) \,\impl\,
F^\ast (\vcc, y)=F^\ast (\vcc, P(\avc));&\label{econd}\\
&(\forall f)(\forall f^*)((\forall y\leq P(\avc)) (F(\vcc,y)=
F(f^*,\vec{x},y))\,\impl &\\
&(\forall y)(F(\vcc,y)= 
F(f^*,\vec{x},y)));&\label{fcond}
\end{eqnarray*}
and that the functional $F(\vcc)$ is defined by 
\begin{equation}
F(\vcc) = F^\ast(\vcc,P(|f|, |\vec{x}|)).
\end{equation}
Then we say that $F(\vcc)$ is defined from functionals $G,H$ 
by {\it polynomially bounded recursion of polynomial length} with bounds 
$(Q,P)$.
 }\end{definition}

\noindent {\bf Theorem}:~~{\sl 
Assume that the functional $F(\vcc)$ 
is defined from the
functionals $G$ and $H$ by polynomially bounded recursion of polynomial 
length with bounds $(Q,P)$. Then the functional $F(\vcc)$ is a 
basic feasible functional.}\\

\subsection{Turing Machine Characterization}

We use the usual model for computability with oracle Turing machines (OTM). 
Function inputs are presented using oracles, which are queried using
separate {\em write-only oracle input tapes} and {\em read-only oracle 
output tapes}, while the machine is in 
the {\em oracle query state} corresponding to the input function
which is queried. To query
function input $f$ at the value $x$, $x$ is written in binary notation
on the oracle input tape associated with $f$, and the corresponding
oracle query state is entered. After entering the oracle state 
which corresponds to $f$, 
the value $f(x)$ appears on the oracle output tape associated with $f$, 
the oracle input tape is then erased and both the writing head of the oracle 
input tape and the reading head of the oracle 
output tape are placed at the corresponding initial cells of the tapes. 
Thus, iterations of
the form $f(f(\ldots f(x) \ldots ))$ cannot be computed without the machine 
having to copy the intermediate results from the oracle output tape to the oracle 
input tape. 
In general, there are two possible conventions 
for accounting for the running time of
an oracle call. In Mehlhorn's model, an oracle call has unit cost,
while in the Kapron/Cook model, the oracle call described above
has a cost of $|f(x)|$ time steps. Mehlhorn \cite{Mehl} and Cook and 
Kapron \cite{KapCook} proved the following theorems.\\
~\\

\noindent{\bf Theorem} (Mehlhorn, \cite{Mehl}):~~{\sl A functional 
$\FF$, where $\vec{f}\in ({\Bbb N}^{\Bbb N})^k$ and 
$\vec{x}\in {\Bbb N}^l$ is basic feasible if and only if there exists
an oracle Turing machine $M$ with oracles for functions $\vec{f}$ 
and a basic feasible functional
$K(\vccc)$ such that $M$ computes $\FF$ and 
the running time  $T(\vccc)$, with a unit cost for each oracle query, 
satisfies
$$(\forall \vec{f})(\forall \vec{x}) (T(\vccc)\leq K(\vccc))$$ }
\noindent{\bf Theorem} (Kapron and Cook, 
\cite{KapCook}):~~{\sl A functional $\FF$, where 
$\vec{f}\in ({\Bbb N}^{\Bbb N})^k$ and 
$\vec{x}\in {\Bbb N}^l$ is basic feasible if and only if there exists
an oracle Turing machine $M$ with oracles for functions $\vec{f}$ 
and a second order polynomial $\PP$ such that $M$ computes $\FF$ and 
the running time $T(\vccc)$, with $|f_i(z)|$ as the cost for an oracle
query of $f_i\in \vec{f}$ at oracle input value $z$, satisfies
$$(\forall \vec{f})(\forall \vec{x}) (T(\vccc)\leq
P(|\vec{f}|,|\vec{x}|))$$ }

\noindent In this paper we improve both theorems by combining their best features. 
\begin{definition}{\sl
A functional $\FF$ is {\it computable in polynomial time} 
if there exists an oracle Turing machine $M$ with oracles for 
functions $\vec{f}$ and a second order polynomial $\PP$ 
such that $M$ computes $\FF$ and for all $\vccc$,
the running time $T(\vccc)$, obtained by counting each oracle 
query as a {\it single step} regardless of the size of the oracle 
output, satisfies
$$T(\vccc)\leq \PP.$$ 
}\end{definition}
\noindent{\bf Theorem}(\ref{pbff}):~~{\sl A functional $\FF$ is a 
polynomial time computable
functional if and only if it is a basic feasible functional.}\\
~\\
\subsection{Formal Theories of Second Order Arithmetic}

Theories ${\bf S}^1_2$ and ${\bf R}^1_2$ 
which we use in this paper
were introduced in \cite{IG1} and used in \cite{CIK}. ${\bf S}^1_2$  is a 
second order extension of Buss's $S^1_2$.
Theory ${\bf S}^1_2$  is formulated in the language 
of $S^1_2$, i.e. $\leq$, 0, 1, +, $\cdot$, $|x|$, $\half{x}$, 
$\#$, extended (for easier bootstrapping) 
by the function $x\res y$ producing
the number consisting of the first, more 
significant $y$ bits of $x$ i.~e.~ $x\res y = \lfloor x/2^{|x|\dotminus 
y}\rfloor$, 
and by a symbol for the application functional $\Ap(f,\vec{x})$.
Consequently, we also extend the usual set of open axioms 
$\BASIC$ of theories of bounded arithmetic 
by adding a few axioms for 
$\res$ (we keep the same notation for the extended set). 

The hierarchy of bounded formulas (i.e.~formulas whose all quantifiers 
are bounded) is obtained from the corresponding 
hierarchies of bounded formulas of the first order bounded arithmetic,
by allowing the application functional $\Ap(f,\vec{x})$ to appear in the
atomic formulas. Bounded quantifiers of the form 
$\exists x \leq |t|$ and $\forall x \leq |t|$
are called {\em sharply bounded} quantifiers.
Bounded formulas form a hierarchy at whose bottom level are {\em sharply 
bounded} or ${\bf \Sigma}^b_0$ formulas, i.e.~formulas 
whose all quantifiers are sharply 
bounded. The complexity of an arbitrary bounded formula (e.g., ${\bf 
\Sigma}^b_i$ , ${\bf \Pi}^b_i$) is obtained by 
counting the alternations of bounded quantifiers ignoring 
sharply bounded ones.

Theory $\SSi$ is obtained from $\BASIC$ by adding either one of the 
following two induction schemas for ${\bf \Sigma}^b_i$ formulas:
\begin{description}
\item [${\bf \Sigma}^b_i\PIND$] 
$\;\;\;\;\; A(0,\vec{f},\vec{y}) \wedge  (\forall x) (A(\half{x},\vec{f},\vec{y})\impl
A(x,\vec{f},\vec{y})) \impl (\forall x) A(x,\vec{f},\vec{y})$
\item [${\bf \Sigma}^b_i\LIND$]
$\;\;\;\;\; A(0,\vec{f},\vec{y}) \wedge  (\forall x) (A(x,\vec{f},\vec{y})\impl
A(x+1,\vec{f},\vec{y})) \impl (\forall x) A(|x|,\vec{f},\vec{y})$
\end{description}
The proof of the fact that theories $(\BASIC + {\bf \Sigma}^b_1\PIND)$ and  
$(\BASIC + {\bf \Sigma}^b_1\LIND)$ are equivalent is not only standard,
(see \cite{Buss}); but is actually facilitated by
the presence of the function $x\res y$ in  our 
language and its basic properties as axioms of $BASIC$.
We will also mention another sequence of formal theories, $\RRi$, 
important for the study of parallel computability of functionals,
obtained from (again slightly extended) 
theory $BASIC$ by adding the following induction schema for ${\bf \Sigma}^b_i$ formulas:

\begin{description}
\item [${\bf \Sigma}^b_i \LPIND$] 
$\;\;\;\;\; A(0,\vec{f},\vec{y}) \wedge  (\forall x) (A(\half{x},\vec{f},\vec{y})\impl
A(x,\vec{f},\vec{y})) \impl (\forall x) A(|x|,\vec{f},\vec{y})$
\end{description}

First order theory $R^1_2$ was studied by Allen (\cite{Allen}) and, 
independently, by Clote and Takeuti (\cite{CT}) ($TNC$ in their 
notation). Our techniques apply equally well 
to parallel computable functionals (a generalization of the class $NC$; see 
\cite{CIK}).

Since we do not have functional substitution in our definition of basic 
feasible functionals, theories $\SSi$ have no comprehension axioms at all. 
The most important property of the above theories is that definable 
function(al)s of these theories correspond to important complexity 
classes of functionals. 
\begin{definition}
A functional $F$ is ${\bf \Sigma}^b_i$ definable in the theory $\SSi$,  
$i \geq 1$, if
there exists a term $t(\vec{f},\vec{x})$ and a ${\bf \Sigma}^b_i$ formula 
$\Psi_F(\vccc,y)$ such that 
$$\SSi\proves\forall \vccc
\,\exists ! y\leq t(\vccc) \,\Psi_F(\vccc,y)$$
and
$$\langle {\Bbb N}^k,({\Bbb N}^{\Bbb N})^m\rangle \models \forall \vccc 
\,\Psi_F(\vccc, F(\vccc)).$$
\end{definition}

For $i\geq 1$ provably ${\bf \Sigma}^b_i$ definable functionals of
$\SSi$  are functionals of the second order polynomial time hierarchy, 
$\bbox^P_i$, a natural extension of the usual polynomial 
time hierarchy for functions $\pbox^P_i$.  In particular, ${\bf \Sigma}^b_1$ 
definable functionals of the theory $\SS$ are exactly functionals
which form the first level of this hierarchy, $\bbox^P_1$ which are 
just the basic feasible functionals. In order to 
prove this fact we use {\em multiple limited recursion on notation} 
(MLRN) due to Cook and Kapron (\cite{KapCook}), of which we give a 
very simple proof. Our argument for $i=1$ easily generalizes to 
$\SSi$ and $\bbox^P_i$ for any $i$.

\subsection{Proof Techniques}

The main novelty of this paper is in the way how we use logic 
(in particular formal theories) as a tool to derive 
results of recursion-theoretic nature as well as results on
Turing computability of functionals. 

It is easy to see that every basic feasible functional $F(x,f)$ is 
computable on a polynomial time 
oracle Turing machine, as well as that it can be obtained using polynomially 
bounded recursion of polynomial length;
this follows immediately from the fact that every basic feasible 
functional can be majorized by a second order polynomial.
The difficult part is to prove that if a functional is polynomial time 
computable or computable from basic feasible functionals 
using polynomially bounded recursion of polynomial 
length, that then it is a basic feasible functional. 

The approach used in the original Cobham's proof for the first order case
is not applicable here, since it is not possible to code directly 
a complete instantaneous description of an oracle Turing machine run
or of the sequence of intermediate values of a recursive procedure 
after $|i|$ many steps and then evaluate this functional at a value of $|i|$ 
which is greater or equal than the number of necessary steps for the 
computation to terminate. This is because both the sizes of complete 
instantaneous descriptions of an oracle Turing machine run
(or of the sequence of intermediate values of a recursive procedure)
and the number of steps necessary for termination of a 
computation of an oracle Turing machine 
(or of a recursive procedure) 
are bounded by second order polynomials which cannot be 
majorized by basic feasible functionals. However, definitions by 
recursion on notation require such bounds. 

We eliminate the need for such bounds by showing that functionals 
computable on a Turing machine or by a recursive procedure with 
bounds which are second order polynomials 
are ${\bf \Sigma}^b_1$ definable in the theory $\SS$, and then, using 
entirely conventional proof-theoretic methods,  
we show that functionals which are ${\bf \Sigma}^b_1$ definable 
in the theory $\SS$ are exactly basic feasible functionals. 

In order to show that functionals defined using second order 
polynomials 
are ${\bf \Sigma}^b_1$ definable in the theory $\SS$ we replace bounds involving
second order polynomials with their suitable representations within our 
formal theory $\SS$. This representation consists of a sequence 
of existentially bounded quantifiers prefixing a term bound, and is based on 
the following Lemma. Without any loss of generality, from now on 
we assume that we have only one input function $f$.\\
~\\
\noindent{\bf Lemma}:~~{\sl
Let $P(\avc)$ be a second order polynomial of depth $d$ (see section 
\ref{preliminaries}); then
there exists a sequence of terms $t_0,\ldots, t_d$ containing only 
operations 
$Ap(f,x), x + y, x \cdot y , x \# y, \half{x}$ and the constant 
$\underline{1}$ such that for all $f$ and all $\vec{x}$ 
the following four formulas are true in the standard model:}
\begin{eqnarray*}
&(\forall u)( |u|\leq \Pp \liff (\exists z_1 \leq t_0(\vec{x})) \ldots 
(\exists z_{d} \leq t_{d-1}(\vcc,z_1,\ldots,z_{d-1}))&\\ 
&(u\leq t_d(\vcc, z_1, \ldots, z_{d})))&
\end{eqnarray*}
Thus, a bound involving a second order polynomial can be replaced by 
a sequence of existential quantifiers bounded by terms. 

We now work in a formal theory and can 
use objects defined by formulas with higher quantifier complexity as 
well. Also, instead of having to provide explicit definitions with 
appropriate bounds, we can use induction and give ``existential'' proofs.
This greatly facilitates our arguments.

\section{Second order polynomials}\label{preliminaries}

The {\it depth} $d(P)$ of a second order polynomial is defined (in 
\cite{KapCook}) to be the maximal number of
nestings of the application functional; thus, the depth of the
polynomials not involving the application functional is equal to $0$ and
$d(P_1 + P_2)=d(P_1 \cdot P_2)=max\{d(P_1),d(P_2)\}$,
while $d(f(P_1))=d(P_1)+1$.

The following simple Lemma can be proved by
induction on the complexity of the definition of
$\BFF$ functionals, using monotonicity of the function $\lambda z.|f|(z)$.
\begin{lemma} (Townsend \cite{Town})
{\sl For every functional $\FF \in \BFF$ 
there exists a second order polynomial $P(|\vec{f}|, |\vec{x}|)$ such that 
for all $\vec{f}$ and for all $\vec{x}$
$$|\FF|\leq
P(|\vec{f}|, |\vec{x}|).$$
}\end{lemma}
In order to handle bounds with second order polynomials using
proof-theoretic means, we inessentially narrow the class of second
order polynomials which we will be using. 
We consider only polynomials which have the property 
that  for every $m$ smaller than the depth of the polynomial $P$ and every 
$f_j$ appearing in $P$,
there exists exactly one sub-polynomial $P^\ast$ of $P$ of the form
$f_j(P^\ast)$ of depth $m$; also, for any two subpolynomials of 
depths $d_{1}$ and $d_{2}$
if $d_1\leq d_2$ then 
for all $\vec{f}$ and all $\vec{x}$, $P_{d_1}(|\vec{f}|,|\vec{x}|) \leq
P_{d_2}(|\vec{f}|,|\vec{x}|)$. Such polynomials are called {\em 
regular} polynomials. Every second order polynomial is majorized by a 
regular second order polynomial. To see this just observe that 
for every function $f$, the function $\lambda z.|f|(z)$ is a monotone increasing
function in $z$ and so we can inductively replace several second order
polynomials which are arguments of functions $|f_j|$ and which are of the
same depth, by their sum. Since we will use second order polynomials only to majorize other
functionals, we will work only with regular second order polynomials.
From now on, when we say that $P$ is a 
second order polynomial, we actually mean that $P$ is a {\it regular} 
second order polynomial. Also, only for the simplicity of our notation, we 
will assume that we have only one second order variable $f$. It is easy to 
see that all our arguments easily generalize to the cases involving 
several second order variables.

The next lemma will be used to replace bounds which are 
second order polynomials with second order 
terms involving only feasible functionals.
We use ``witnessing points'' to approximate from below 
the value of a second order polynomial. 

\begin{lemma}\label{d}{\sl
Let $P(\avc)$ be a second order polynomial of depth $d$; then
there exists a sequence of terms $t_0,\ldots, t_d$ containing only 
operations $Ap(f,x), x + y, x \cdot y , x \# y, \half{x}$ and the constant 
$\underline{1}$ such that for all $f$, all $\vec{x}$ and all $u$:}
\begin{eqnarray*}
&	|u|\leq \Pp\; \liff\; (\exists z_1 \leq t_0(\vec{x})) 
(\exists z_2 \leq t_1(\vcc,z_1))\ldots &\\ 
&\,\,\,\,\,\,\,\,\,\,\,\,\,\;\;\;\;\;\;\;\;\;\;\;\;\;\;\;\;\;\;
(\exists z_{d} \leq t_{d-1}(\vcc,z_1,\ldots,z_{d-1})) 
(u\leq t_d(\vcc, z_1, \ldots, z_{d}))&
\end{eqnarray*}
\end{lemma}
\PF Simple.
\begin{definition} {\sl 
Let $P(\avc)$ be a second order polynomial and
$t_0(\vec{x})$,$\ldots$,\\$t_d(\vcc,z_1,\ldots,z_d)$ be the sequence of terms as 
in Lemma~\paref{d}; then we call this sequence {\it the sequence of terms 
associated with the polynomial} $P(\avc)$. We will often abbreviate
the sequence of these terms as $\vec{t^P}$, and the corresponding
quantifier prefix 
$(\exists z_1 \leq t_0(\vec{x})) (\exists z_2 \leq t_1(\vcc,z_1))\ldots 
(\exists z_{d} \leq t_{d-1}(\vcc,z_1,\ldots,z_{d-1}))$
as $(\exists \vec{z}\leq \vec{t^P})$.
}\end{definition}

\section{Theories of Second Order Arithmetic}\label{theories}

We now define two second order formal
theories of arithmetic ${\bf S}^1_2$ and ${\bf R}^1_2$; one sort
of variables range over the set of natural numbers $\Bbb N$; 
the other range over the set of functions of
type ${\Bbb N}\rightarrow {\Bbb N}$. We put no constraints on the
growth rate of functions. These two theories are then used to characterize
type 2 feasible functionals in the same way how Buss's $S^1_2$ is used
to characterize feasible functions.

We will use Buss's results on
introducing the polynomial time computable functions in $S^1_2$ (see 
\cite{Buss}), but
due to the presence of functions which can be of an arbitrary growth
rate, we must do more work. Also, to facilitate the bootstrapping of
our theories, we will add a few more polynomial time functions to the
language of our theories which are second order versions of Buss's $S^1_2$.

\begin{definition} {\sl
We will denote by $L_b^2$ the language consisting of the following 
symbols
$\leq,$ 0, 1, +,$\cdot$, $|x|$, $\half{x}$ , $\#$, $\res$ and $\Ap(f,\vec{x})$.
Here $x \res y$ denotes\footnote{Notice that $x\res y$ is $MSP(x,y)$ 
in Buss's notation which we here conveniently
simplify.} the $y$ most significant
bits  of $x$. $\Ap(f,\vec{x})$ is the only mixed-sort symbol for the
application functional whose value is $f(\vec{x})$ in the standard
interpretation.} 
\end{definition}

We will consider the following induction schemas.
\begin{description}

\item [$\Sigma^b_1$-PIND] 
$A(0,\vec{\alpha}) \wedge  (\forall x) (A(\half{x},\vec{\alpha})\impl
A(x)) \impl (\forall x) A(x,\vec{\alpha})$

\item [$\Sigma^b_1$-LIND] 
$A(0,\vec{\alpha}) \wedge  (\forall x) (A(x,\vec{\alpha})\impl A(x+1,\vec{\alpha}))
\impl (\forall x)A(|x|,\vec{\alpha})$
\item [$\Sigma^b_i-LPIND$] 
$A(0,\vec{\alpha}) \wedge  (\forall x) (A(\half{x},\vec{\alpha})\impl
A(x,\vec{\alpha})) \impl (\forall x) A(|x|,\vec{\alpha})$
\item [$\Sigma^b_i-LLIND$] 
$A(0,\vec{\alpha}) \wedge  (\forall x) (A(x,\vec{\alpha})\impl A(x+1,\vec{\alpha}))
\impl (\forall x)A(||x||,\vec{\alpha})$
\end{description}
One can easily check that by adding to $\BASIC$ the following axioms for 
$x\res y$: $x\res 0=0$, $x \geq 1 \; \rightarrow \;  x\res 1=1$, $y < |x| \;\rightarrow \;
x\res y = \fr{x\res (y+1)}{2}$ and 
$y \geq |x| \; \rightarrow \;  x\res y = x$ 
makes the usual proof (see Buss's \cite{Buss}) of
$\BASIC+({\bf \Sigma}^b_i\LIND)\; \equiv\; \BASIC+({\bf \Sigma}^b_i \PIND)$ simple, 
and  that it goes through also for $i=1$ equally easily. 
Theories $\SSi$ are obtained by adding to such slightly extended 
$\BASIC$ either $({\bf \Sigma}^b_i\LIND)$ or $({\bf \Sigma}^b_i \PIND)$.
Theories ${\bf R}^i_2$ are obtained by first extending the language 
$L^{2}_{b}$ into the language $L^{2}_{d} = L^{2}_{b}\cup \{
\dotminus, Bit(x,y), $Lsp(x,y) \}; then by expanding the axioms of 
$\BASIC$ into $\BASIC^{+}$ as described in \cite{Allen}, by adding 
basic properties of these functions and finnaly adding either of the 
two schemas $\Sigma^b_i-LPIND$ or $\Sigma^b_i-LLIND$. 

We now concentrate on theory $\BS$ as our main tool for studying the 
properties of basic feasible functionals. Let 
$\vec{\alpha}=\vec{f},\vec{x}$. We now want to show that 
a functional is a basic feasible functional if and only if it is ${\bf \Sigma}^b_1$ 
definable in the theory $\BS$. We start with the easier direction.

\begin{theorem}{\sl
Every basic feasible functional is ${\bf \Sigma}^b_1$ definable in the 
theory}\\ $\BS$.
\end{theorem}
\PF
We prove by induction on complexity of definition of a functional $F$ 
that there exists a ${\bf \Sigma}^b_1$ formula
$\Theta_F(\vec{\alpha},\vec{z}_1, \ldots, \vec{z}_k,y)$
and sequences of terms $\vec{t}_1, \ldots, \vec{t}_k, t_{k+1}$
such that
\begin{eqnarray}\label{graph}
\lefteqn{\BS  \vdash \forall \vec{\alpha}
\,\exists \vec{z}_1 \le \vec{t}_1(\vec{\alpha}) \ldots
\exists \vec{z}_k \le \vec{t}_k(\vec{\alpha},\vec{z}_1, \ldots, \vec{z}_{k-1})}\\
& & \hspace*{1in}  \exists ! y \le 
t_{k+1}(\vec{\alpha},\vec{z}_1, \ldots, \vec{z}_k)
\,\Theta_F(\vec{\alpha},\vec{z}_1, \ldots, \vec{z}_k, y)\nonumber
\end{eqnarray}
which clearly implies that $F$ is $\Sigma^b_1$ definable in $\BS$. The 
proof proceeds by induction on the definition of 
$F \in \BFF$. If $F$ is obtained by functional
composition or by expansion the proof is straightforward.

If $F$ is defined by limited recursion on
notation from $G,H$ and $K$, we use Buss's function 
$\SqBd(a,b) = (2b+1)\#(4(2a+1)^2)$ (see \cite{Buss}),
which puts an upper bound on codes of sequences of
length at most $|b|+1$, consisting of numbers $\le a$. Then, assuming that
\begin{eqnarray*}
F(\vec{\alpha},0) & = & G(\vec{\alpha}) \\
F(\vec{\alpha},y) & = & H(\vec{\alpha},y,F(\vec{\alpha},\half{y})) \\
F(\vec{\alpha},y) & \le & K(\vec{\alpha},y)
\end{eqnarray*}
by induction hypothesis
there are formulas $\Theta_{H}, \Theta_{G}, \Theta_{K}$
and terms $\vec{t_i}^{H}, \vec{t_j}^{G}, \vec{t_r}^{K},$
$1 \le i,j,r \le k+1$
such that
\begin{eqnarray}\label{(1)}
\BS  &\vdash& \forall \vec{\alpha}\,\prefixg \,
\exists g \le t_{k+1}^{\,G}(\vec{\alpha},\vec{\bf{z}}^{\,G})
\,\Theta_{G}(\vec{\alpha},\vec{\bf{z}}^{\,G},g),\nonumber\\
\BS  &\vdash &\forall \vec{\alpha} \,\forall y \,\forall v
\,\prefixH\,\exists h \le t_{k+1}^H(\vec{\alpha},y,v,\vec{\bf{z}}^H)\, 
\Theta_H(\vec{\alpha},y,v,\vec{\bf{z}}^H,h)\nonumber\\
\BS &\vdash & \forall \vec{\alpha} \,\forall y
\,\prefixK\,\exists q \le t_{k+1}^K(\vec{\alpha},y,\vec{\bf{z}}^K)\, 
\Theta_K(\vec{\alpha},y,\vec{\bf{z}}^K,q).\nonumber\\
\end{eqnarray}
then we prove
\begin{eqnarray*}
&\BS  \vdash \forall \vec{\alpha} \,\forall y \,\prefixK \,\exists t \le y 
\,\exists w \le \SqBd(t_{k+1}^K (\vec{\alpha},t,\vec{\bf{z}}^K),y)&\\
&\exists z \le w \Theta_F(\vec{\alpha},y,t,w,\vec{\bf{z}}^K,z) &
\end{eqnarray*}
where $\Theta_F$ is the conjunction of the following formulas:
\begin{eqnarray*} 
&\prefixg\,\exists g \le t_{k+1}^{\,G} (\vec{\alpha},\vec{\bf{z}}^{\,G})
\,\,(\Theta_{G}(\vec{\alpha},\vec{\bf{z}}^{\,G},g) \,\wedge 
\exists q_0 \le t_{k+1}^{K} (\vec{\alpha}, t,\vec{\bf{z}}^K)&\\
&\Theta_{K}(\vec{\alpha},0,\vec{\bf{z}}^{K},q_0) \,\wedge 
 (w)_0 = min(g,q_0))  &
\end{eqnarray*}
\begin{eqnarray*}
&(\forall i < |y| \,\exists q_{i+1} \le t_{k+1}^{K} 
(\vec{\alpha}, t,\vec{\bf{z}}^K)
\,\,\Theta_{K}(\vec{\alpha},y \res 
(i+1),\vec{\bf{z}}^{K},q_{i+1})&\\
&\wedge (w)_i \leq t_{k+1}^{K} (\vec{\alpha}, t,\vec{\bf{z}}^K) \,\wedge \prefixH \,\exists h \le t_{k+1}^{H}(\vec{\alpha},y \res i, (w)_i, 
\vec{\bf{z}}^H)\,\,&\\
&\Theta_{H}(\vec{\alpha},y \res i, (w)_i,\vec{\bf{z}}^{H},h) 
\,\wedge (w)_{i+1} = min(h,q_{i+1})) \wedge ((w)_{|y|} = z)&
\end{eqnarray*}
\begin{eqnarray*}
\left(\forall j \le |y|
\,\exists q \le t_{k+1}^{K} (\vec{\alpha}, t,\vec{\bf{z}}^K)
\,\,\Theta_{K}(\vec{\alpha},y \res j,\vec{\bf{z}}^{K},q)\right)
\end{eqnarray*}
The proof is straightforward; the first formula corresponds the initial value of 
the function the second to the recursion on  notation and the third to  
bounding and the final value of the computation.

To prove the converse we need definitions by {\it multiple limited 
recursion on notation} MLRN due to Cook and Kapron, of which we give 
a very simple proof. 

\begin{theorem}\label{mix}{\sl
Let $G_1, G_2, H_1, H_2, K_1$  and $K_2$ be basic feasible functionals, 
and assume that $F_1$ and $F_2$ satisfy 
\begin{eqnarray}\label{multa}
F_1(0,\vec{\alpha}) & = & G_1(\vec{\alpha})\nonumber\\
F_2(0,\vec{\alpha}) & = & G_2(\vec{\alpha})\nonumber\\
F_1(u,\vec{\alpha}) & = & 
H_1(u,F_1(\half{u},\vec{\alpha}),F_2(\half{u},\vec{\alpha}),
\vec{\alpha})\\
F_2(u,\vec{\alpha}) & = & 
H_2(u,F_1(\half{u},\vec{\alpha}),F_2(\half{u},\vec{\alpha}),
\vec{\alpha})\nonumber\\
F_1(u,\vec{\alpha}) & \leq & K_1(u,\vec{\alpha})\nonumber\\
F_2(u,\vec{\alpha}) & \leq & 
K_2(u,\vec{\alpha},F_1(u,\vec{\alpha}))\nonumber
\end{eqnarray}
Then $F_1$ and  $F_2$ are also basic feasible functionals.}
\end{theorem}
\PF 
Let $\hat{K}$ be defined as follows:
$$\hat{K}(0,\vec{\alpha})  = 0$$
\[ \hat{K}(u,\vec{\alpha}) = \left\{ \begin{array}{ll}
\hat{K}(\half{u},\vec{\alpha}) & \mbox{if $K_1
(u,\vec{\alpha}) \le K_1(\hat{K}(\half{u},\vec{\alpha}),\vec{\alpha})$} 
\nonumber\\
u & \mbox{otherwise}
\end{array}
\right. \]
$$\hat{K}(u,\vec{\alpha})\leq u$$
and let $\bar{K}(u,\vec{\alpha}) = 
K_1(\hat{K}(u,\vec{\alpha}),\vec{\alpha})$. Thus, as it is easily proved 
by induction on $u$, $\bar{K}(u,\vec{\alpha})$ is the largest of the value 
of 
$K_1(u\res i,\vec{\alpha})$, for $u$ fixed and $0\leq i \leq |u|$. Thus, 
the function 
$$S(u,\vec{\alpha})= SqBd(\bar{K}(u,\vec{\alpha}),u)$$ 
has the property 
that it bounds any sequence $a$ of length $|u|$ satisfying: 
$(\forall i\leq |u|)((a)_i\leq K_1(u\res i,\vec{\alpha}))$.

Consider now functionals $F$ and $W$ defined as follows:
\begin{eqnarray}\label{newa}
F(0,\vec{\alpha},w) & = & G_2(\vec{\alpha})\nonumber\\
F(u,\vec{\alpha},w) 
& = & min\{H_2(u,(w)_{(|u|\dotminus 
1)},F(\half{u},\vec{\alpha},w),\vec{\alpha}),\ 
K_2(u,\vec{\alpha},(w)_{|u|})\}\nonumber
\end{eqnarray}
and 
\begin{eqnarray}\label{newb}
&W(0,\vec{\alpha})  =  \left<G_1(\vec{\alpha})\right>&\nonumber\\
&W(u,\vec{\alpha})  =  min\{W(\half{u},\vec{\alpha})^\frown 
H_1 (u,(W(\half{u},\vec{\alpha}))_{|\half{u}|},&\\
& F(\half{u},\vec{\alpha},W(\half{u},\vec{\alpha})),\vec{\alpha}),\ 
S(u,\vec{\alpha})\}&\nonumber
\end{eqnarray}
where $w^\frown a$ stands for the sequence $w$ extended at the end 
with an extra term. Then, clearly, 
\begin{eqnarray}
F(u,\vec{\alpha},w) & \leq & K_2(u,\vec{\alpha},(w)_{|u|})\\
W(u,\vec{\alpha}) & \leq & S(u,\vec{\alpha})
\end{eqnarray}
which implies that these functionals are both basic feasible.
Let now
\begin{eqnarray}
F_1(u,\vec{\alpha}) &= & (W(u,\vec{\alpha}))_{|u|}
	\label{nf} \\
F_2(u,\vec{\alpha}) & = & F(u,\vec{\alpha},W(u,\vec{\alpha}))
	\label{nnff}
\end{eqnarray}
then one can easily prove by (polynomial) induction on $u$ that 
these functionals satisfy the recursive schema (\ref{multa}).
The theorem easily generalizes to $n > 2$ functionals, in which case 
we proceed by induction and assume that the claim of the Lemma holds for $n-1$. 
Replace the first functional, $F_1$, by a variable
$w$, representing the code of the computation of a functional $W$, 
as we did in the case of $n = 2$; hence the number of the remaining 
functionals is $n-1$.
\qed

We now want to show that every functional of type 2 which is 
${\bf \Sigma}^b_1$ definable in $\BS$ 
is a basic feasible functional.
The proof of this fact is identical to the Buss proof of the 
coresponding result for $S^{1}_{2}$ and the polynomial time computable 
functions, with one single exception. For convenience, instead of the 
original Buss's proof, we will use Sieg's method of Herbrand Analyses 
from \cite{Sieg1}.  The following Lemma replaces Lemma 1.3.4. 
from \cite{Sieg2}. The rest of the proof is identical to the first 
order case; the fact
that we have a two sorted language has no impact on the rest of the proof. 

Exactly as in the first order case, we introduce auxiliary theories needed 
for the proof.\\ 
$\PVF$ stands for the theory on a language whith a symbol for every 
functional in $\BFFM$. Besides the basic axioms it also has defining 
recursion equations for every such symbol, and induction 
schema for every open formula.

$(n-{\bf \Sigma}^b_1)-PIND$ stands for the theory which 
extends $\PVF$ with the induction 
schema for formulas in prenex normal form whose prefix has at most 
$n$ bounded existential quantifiers and no sharply bounded universal 
quantifiers. ${\bf \Sigma}^b_1$ formulas without any sharply bounded 
universal quantifiers are called {\em strict} ${\bf \Sigma}^b_1$ formulas.
\begin{lemma}\label{BS_PVF}
Let $\Delta$ be a set containing only existential formulas
(with the existential quantifier bounded or unbounded). Then if
$(n-{\bf \Sigma}^b_1)-PIND\proves\Delta$, then also $\PVF\proves\Delta$.
\end{lemma}
\PF We proceed by induction on the number of applications of the induction 
rule applied to $(n-{\bf \Sigma^b_1})$-formulas which can contain
at most n existential quantifiers (i.e.~we {\em do not} count instances of
induction rule applied to open formulas) in the I-normal
derivations in $(n-{\bf \Sigma}^b_1)-PIND$ of sets of 
$(n-{\bf \Sigma^b_1})$-formulas. (Recall that a derivation is I-normal 
if all cuts are on induction or atomic formulas only.)

Let the claim of the Lemma hold for derivations with $k$ applications 
of the induction rule.  Assume that ${(n-\bf \Sigma^b_1})-PIND\proves\Delta$ with an 
I-normal derivation $d$ with $k+1$ such applications of the induction rule; 
consider a {\em top-most} instance of the induction rule applied to a 
${(n-\bf \Sigma^b_1})$-formula $\psi$ (i.e.~an application of the 
induction rule so that all other applications of the induction rule 
appearing above it in the derivation $d$ are on open formulas):
$$\pind{}{\psi}{t}.$$

We want to prove that we can reduce this application of 
$(n-\bf{\Sigma^b_1})-PIND$
to an application of open induction.

If $n = 2$, then $\psi\left(b_i,\,\vec{\alpha}\right)$ is of the 
form 
$$\exists z_1 \le t_0\left(b_i,\,\vec{\alpha}\right) \,\exists z_2 
\le t_1\left(b_i,\,\vec{\alpha},\,z_1\right) 
\,\theta\left(b_i,\,z_1,\,z_2,\,\vec{\alpha}\right),$$ where $b_i$ 
does not appear in the sequence
of variables $\vec{\alpha}$, and $\theta$ is open. 

Let $d_1$ and $d_2$ be the immediate subderivations; 
by our assumption, they are derivations in the theory $\PVF$.
We first replace the derivation $d_2$
leading to the set 
$\Gamma(\vec{\alpha}), \neg 
\psi\left(\half{b_i},\,\vec{\alpha}\right), 
\psi\left(b_i,\,\vec{\alpha}\right)$
by a derivation $d_2'$ of the set 
$\Gamma(\vec{\alpha}),  
\,a\leq t_0\left(\half{b_i},\,\vec{\alpha}\right) \,\wedge
\,c\leq t_1\left(\half{b_i},\,\vec{\alpha},a\right) \impl 
\neg\theta\left(\half{b_i},\,a,\,c,\,\vec{\alpha}\right), 
\exists z_1 \leq t_0\left(b_i,\,\vec{\alpha}\right)\, 
\exists z_2 \le t_1\left(b_i,\,\vec{\alpha},\,z_1\right) \,\theta 
\left(b_i,\,z_1,\,z_2,\,\vec{\alpha}\right),$
where $a$ and $c$ are new variables not previously used in $d$.
As in the first order case, we can find functionals 
$\bar{F}_0$, $\hat{F}_0$, $\bar{F}_1$ and $\hat{F}_1$ and
derivations $d^*_1, d^*_2$ of the sets
\begin{eqnarray}\label{0}
\Gamma(\vec{\alpha}), 
\bar{F}_0\left(\vec{\alpha}\right)\le 
t_0\left(0,\,\vec{\alpha}\right)\,\wedge 
\,\hat{F}_0\left(\vec{\alpha}\right)\le 
t_1\left(0,\,\vec{\alpha},\,\bar{F}_0(\vec{\alpha})\right)\,\wedge 
\,\theta\left(0,\,\bar{F}_0(\vec{\alpha}),\,\hat{F}_0(\vec{\alpha}),\,\vec{\alpha}\right)\nonumber
\end{eqnarray}
and
\begin{eqnarray}\label{1}
&\Gamma(\vec{\alpha}), 
a\leq t_0\left(\half{b_i},\,\vec{\alpha}\right) \,\wedge 
\,c\leq t_1\left(\half{b_i},\,\vec{\alpha},\,a\right)
\impl 
\neg\theta\left(\half{b_i},\,a,\,c,\,\vec{\alpha}\right),&\nonumber\\
&\bar{F}_1\left(b_i,\,a,\,c,\,\vec{\alpha}\right)\leq 
t_0\left(b_i,\,\vec{\alpha}\right)\, \wedge \hat{F}_1\left(b_i,\,a,\,c,\,\vec{\alpha}\right)\leq 
t_1\left(b_i,\,\vec{\alpha},\,\bar{F}_1(b_i,a,c,\vec{\alpha})\right)&\nonumber \\
&\wedge
\,\theta\left(b_i,\,
\bar{F}_1( 
b_i,a,c,\vec{\alpha}),\,\hat{F}_1(b_i,a,c,\vec{\alpha}),\,
\vec{\alpha}\right).&\nonumber
\end{eqnarray}
Consider now the formula 
\begin{eqnarray*}
&\phi\left(b_i,\,x,\,y,\,\vec{\alpha}\right)\;\; \equiv \;\;x 
\leq t_0\left(b_i,\,\vec{\alpha}\right)\wedge&\\
&y \leq t_1\left(b_i,\,\vec{\alpha},\,x\right)\wedge 
\theta\left(b_i,\,x,\,y,\,\vec{\alpha}\right);&	
\end{eqnarray*}
then the above sets are of the form
$$\Gamma(\vec{\alpha}),\,
\phi\left(0,\,\bar{F}_0(\vec{\alpha}),\,
\hat{F}_0(\vec{\alpha}),\,\vec{\alpha}\right)$$
and
$$\Gamma(\vec{\alpha}), 
\,\neg\phi\left(\half{b_i},\,a,\,c,\,\vec{\alpha}\right),
\phi\left(b_i,\,\bar{F}_1(b_i,a,c,\vec{\alpha}),\,
\hat{F}_1(b_i,a,c,\vec{\alpha}),\,\vec{\alpha}\right)$$
respectively. Thus,
in order to be able to apply the induction rule for open formulas, we
must find two functionals $\bar{F}$ and $\hat{F}$ such that after 
substituting the free variables $a$ and $c$ with the functionals 
$\bar{F}(b_i,\vec{\alpha})$ and $\hat{F}(b_i,\vec{\alpha}),$
respectively, in the following formulas:
$$\phi\left(0,\bar{F}_0(\vec{\alpha}), \hat{F}_0(\vec{\alpha}),
\vec{\alpha}\right),\,\,\,\neg\phi\left(\half{b_i},a,c,\vec{\alpha}\right),\,\,\,
\phi\left(b_i,\bar{F}_1(b_i,a,c,\vec{\alpha}),\hat{F}_1(b_i,a,c,
\vec{\alpha}),\vec{\alpha}\right)$$
they become of the form 
$\sigma(0,\vec{\alpha}),\;\;\sigma(\half{b_i},\vec{\alpha})$ 
and $\sigma(b_i,\vec{\alpha})$ 
respectively.
This suggests the following definitions of $\bar{F}$ and $\hat{F}$: 
\begin{eqnarray}\label{old}
\bar{F}\left(0,\vec{\alpha}\right) & = & 
\bar{F}_0\left(\vec{\alpha}\right)\nonumber\\
\bar{F}\left(b_i,\vec{\alpha}\right) & = & 
\bar{F}_1\left(b_i,\bar{F}(\half{b_i},\vec{\alpha}),\hat{F}
(\half{b_i},\vec{\alpha}),\vec{\alpha}\right)\nonumber\\
\bar{F}\left(b_i,\vec{\alpha}\right) & \leq & 
t_0\left(b_i,\vec{\alpha}\right)\nonumber\\
{\rm and} \nonumber\\
\hat{F}\left(0,\vec{\alpha}\right) & = & 
\hat{F}_0\left(\vec{\alpha}\right)\nonumber\\
\hat{F}\left(b_i,\vec{\alpha}\right) & = & 
\hat{F}_1\left(b_i,\bar{F}(\half{b_i},\vec{\alpha}),\hat{F}
(\half{b_i},\vec{\alpha}),\vec{\alpha}\right)\nonumber\\
\hat{F}\left(b_i,\vec{\alpha}\right) & \leq & 
t_1\left(b_i,\vec{\alpha},\bar{F}(b_i,\vec{\alpha})\right)\nonumber
\end{eqnarray}
By MLRN such definition is correct.

Now we can apply the induction rule for open formulas and 
bounded existential introduction rule to get a $\PVF$ derivation of 
$\Gamma(\vec{\alpha}),\psi(t(\vec{\alpha}),\vec{\alpha}).$
The claim follows now from the inductive hypothesis.

\begin{theorem}
The class of provably total functions of $\BS$ is exactly the class of the 
$\BFFM$ functionals.
\end{theorem}
\PF 

As in the first order case one first 
shows that if $\BS\proves (\forall \vec{\alpha})\exists y\leq 
t(\vec{\alpha}) 
\phi(\vec{\alpha},y)$ where $\phi$ is a ${\bf \Sigma^b_1}$ 
formula, then for some $n$, $(n-{\bf \Sigma^b_1})-PIND\proves (\forall \vec{\alpha})(\exists y\leq 
t(\vec{\alpha})) 
\phi(\vec{\alpha},y)$ and we can also assume that $\phi$ is a strict ${\bf \Sigma^b_1}$ 
formula of the language which includes a symbol for every function 
from $\BFFM$. Then by the previous Lemma $\PVF\proves (\forall \vec{\alpha})(\exists y\leq 
t(\vec{\alpha}) ) 
\phi(\vec{\alpha},y)$. Now the claim of the Theorem
follows from the corfresponding fact about $\PVF$ (whose proof is 
identical to the first order case).
\qed

The following lemma is a crucial tool in our proofs. 
\begin{lemma}\label{PB} {\sl
Let $P(\avc)$ be a second order polynomial of depth $p$ and 
$\vec{t^P} = t^P_0(\vec{x}),\ldots,t^P_p(\vcc,z_1,\ldots,z_p)$ the 
corresponding sequence of terms in variables $\vec{z}^P = 
z_1^P,\ldots,z_p^P$ 
associated with the polynomial $P$. Then
\begin{eqnarray*}\label{mmm}
&{\bf R}^1_2\proves (\forall f)(\forall \vec{x})(\exists u)
(\forall z_1 \leq t^P_0(\vec{x}))\ldots 
(\forall z_{p} \leq t^P_{p-1}(\vcc,z_1,\ldots,z_{p-1}))&\\
&(|t^P_p(\vcc,z_1,\ldots,z_p)|\leq |u|)&
\end{eqnarray*}
}\end{lemma}

Thus, despite the fact that second order polynomials are not basic 
feasible, they are provably bounded even in the theory $\RR$. Notice that 
the reason why, of course, we cannot conclude that second order polynomials 
are feasible functionals from the above provability layes in the fact 
that formula which defines the graph of the functional
is not a ${\bf \Sigma}^b_1$ formula but a
conjunction of a ${\bf \Sigma}^b_1$ and a  ${\bf \Pi}^b_1$ formula:
\begin{eqnarray*}
&\left((\forall z_1 \leq t^P_0(\vec{x}))\ldots 
(\forall z_{p} \leq t^P_{p_1}(\vcc,z_1,\ldots,z_{p_1}))
(t^P_p(\vcc,z_1,\ldots,z_p)\leq u )\right) \wedge &\nonumber\\
&\left((\exists z_1 \leq t_0(\vec{x}))\ldots 
(\exists z_{d} \leq t_{d_1}(\vcc,z_1,\ldots,z_{d_1})) 
(t_d(\vcc, z_1, \ldots, z_{d})= u ))\right)&\nonumber
\end{eqnarray*}
\PF Assume the opposite and fix $\vcc$ such that for the formula 
\begin{eqnarray}\label{unb}
&\Psi(v,\vcc)\equiv
(\exists z_1 \leq t^P_0(\vec{x}))\ldots 
(\exists z_{p} \leq t^P_{p-1}(\vcc,z_1,\ldots,z_{p-1}))&\nonumber\\
&(|t^P_p(\vcc,z_1,\ldots,z_p)|\geq v)&
\end{eqnarray}
we have $(\forall u)\Psi(|u|,\vcc)$.
Fix now an arbitrary $v$ and consider the formula $\Psi^*(i,v,\vcc) 
\equiv \Psi(v\res i, \vcc)$. Then clearly $\Psi^*(0,v,\vcc)$
holds. Assume now that $\Psi^*(i,v,\vcc)$ holds; then for some 
$\underline{z}_1,\ldots,\underline{z}_{p-1}$ we have 
$(|t^P_p(\vcc,\underline{z}_1,\ldots,\underline{z}_p)|\geq v\res i).$
Let $c_i = t^P_p(\vcc,\underline{z}_1,\ldots,\underline{z}_p)$; then
$v\res i \leq c_i$. But then $|v\res (2i)| \leq 2|v\res i| = |(v\res 
i)^2|$ 
and consequently $v\res(2i) \leq 2(v\res i)^2 \leq 2|c|^2 = |(c \# c)^2|$.
Thus, $(\exists d\leq (c \# c)^2)(v\res (2i)=|d|)$; since
$(\forall u)\Psi(|u|,\vcc)$ we get $\Psi(v\res (2i),\vcc)$,
i.e.~$\Psi^*(2i,v,\vcc)$. The above argument shows that for the
${\bf \Sigma}^b_1$ formula $\Psi^*$ the following holds:
$$\Psi^*(0,v,\vcc) \; \wedge (\forall i)(\Psi^*(i,v,\vcc)
\impl \Psi^*(2i,v,\vcc)).$$
Using the corresponding instance of the ${\bf \Sigma}^b_1$-LPIND, we get
$(\forall i)\Psi^*(v,|i|,\vcc)$. Thus,
$\Psi^*(v\res|v|,\vcc)$, i.e.~$\Psi(v,\vcc)$. Consequently,
$(\exists z_1 \leq t^P_0(\vec{x}))\ldots 
(\exists z_{p} \leq t^P_{p-1}(\vcc,z_1,\ldots,z_{p-1}))
(|t^P_p(\vcc,z_1,\ldots,z_p)|\geq v)$
holds; since $v$ was an arbitrary parameter, we get 
\begin{equation}\label{exp}
(\forall v)(\exists w)(v = |w|).
\end{equation} 
Property \paref{exp} and ${\bf \Sigma}^b_1$-LPIND imply
that ${\bf \Sigma}^b_1$-PIND and consequently also ${\bf \Sigma}^b_1$-LIND both
hold. (In fact, using \paref{exp} again, we can get that ${\bf \Sigma}^b_1$-IND 
and
$\Pi^b_1$-IND also hold.) We now show that for all $k\leq p$ 
the following holds: 
\begin{equation}\label{cont}
(\exists u_k)(\forall z_1 \leq t^P_0(\vec{x}))\ldots 
(\forall z_{k} \leq t^P_{k-1}(\vcc,z_1,\ldots,z_{k-1}))
(t^P_k(\vcc,z_1,\ldots,z_k)\leq u_k)
\end{equation}
For $k=0$, using \paref{exp}, we find $w_0$ such that 
$t^P_0(\vec{x})=|w_0|$ and then apply $\Pi^b_1$-LIND on $y$ in the formula
\begin{equation}\label{gggg}
(\exists z^*_0\leq |w_0|)(\forall z_0\leq y)
(t^P_1(\vcc,z_0)\leq t^P_1(\vcc,z^*_0))
\end{equation}
to get
$(\exists z^*_0\leq |w_0|)(\forall z_0\leq |w_0|)
(t^P_1(\vcc,z_0)\leq t^P_1(\vcc,z^*_0))$, i.e.~
$(\exists z^*_0\leq t^P_0(\vec{x}))(\forall z_0\leq t^P(\vec{x}))
(t^P_1(\vcc,z_0)\leq t^P_1(\vcc,z^*_0))$. We now take 
$u_0 = t^P_1(\vcc,z^*_0)$.

Similarly, assuming that 
$$(\exists u_k)(\forall z_1 \leq t^P_0(\vec{x}))\ldots 
(\forall z_{k} \leq t^P_{k-1}(\vcc,z_1,\ldots,z_{k-1}))
(t^P_k(\vcc,z_1,\ldots,z_k) \leq u_k)$$ and taking $w_k$ such that 
$|w_k| = u_k$ 
we prove by induction 
$$(\exists z^*_{k+1}\leq |w_k|)(\forall z_{k+1}\leq |w_k|)
(t^P_{k+1}(\vcc, z_{k+1})\leq t^P_{k+1}(\vcc, z^*_{k+1})).$$
Taking $u = t^P_{k+1}(\vcc, z^*_{k+1})$ we get a contradiction 
with \paref{mmm}.\qed

We can now prove our main results.

\begin{theorem}
Assume that the functional $F(\vcc,y)$ is defined from the functionals 
$G(\vcc)$ and $H(\vcc,z,y)$ by polynomially bounded
recursion on notation with the bound $Q(|f|, |\vec{x}|,|y|)$. Then
$F(\vcc,y)$ is a basic feasible functional.
\end{theorem}
\PF 
Just replace second order polynomial bound with an appropriate ${\bf \Sigma}^b_1$ formula.
Then a simple application of ${\bf \Sigma}^b_1-LIND$ proves that such 
a ${\bf \Sigma}^b_1$ formula defines a total function. 
\qed 

\begin{theorem}
Assume that the functional $F(\vcc)$ 
is defined from the
functionals $G$ and $H$ by polynomially bounded recursion of polynomial 
length with bounds $(Q,P)$. Then the functional $F(\vcc)$ is a 
basic feasible functionals.
\end{theorem}
\PF 
Assume that $P(|f|, |\vec{x}|)$ and $Q(\avc)$ are two
second order polynomials and let 
$G(\vcc)$ and  $H(\vcc,y,z)$
be two basic feasible functionals. Let the functional   
$F^\ast(\vcc,y)$ satisfies 
\begin{eqnarray*}
&F^\ast(\vcc,0) = G(\vcc)&\\
&F^\ast(\vcc,y+1) = H(\vcc,F^\ast(\vcc,y),y),&\\
&(\forall y \leq P(|f|, |\vec{x}|))(|F^\ast (\vcc, y)|\leq Q(|f|, 
|\vec{x}|))&\nonumber\\
&(\forall y) (y \geq P(\avc) \,\impl\,
F^\ast (\vcc, y)=F^\ast (\vcc, P(\avc))&\nonumber\\
&(\forall f)(\forall f^*)((\forall y\leq P(\avc)) (F^\ast(\vcc,y)=
F^\ast(f^*,\vec{x},y))\,\impl&\\
&(\forall y)(F^\ast(\vcc,y)= 
F^\ast(f^*,\vec{x},y)))&
\end{eqnarray*}
and that the functional $F(\vcc)$ is defined by 
\begin{equation}
F(\vcc) = F^\ast(\vcc,P(|f|, |\vec{x}|)).\nonumber
\end{equation}
We replace the functional $F(\vcc)$ by $\hat{F}(\vcc)$ defined as 
follows. At each recursive stage $u+1$ of the computation of 
$F^\ast(\vcc,u+1)$ we also evaluate an approximation 
$P^{*}(|f|, |\vec{x}|,u)$ of $P(|f|, |\vec{x}|)$ on the basis of the 
values $f(z_{i})$ which were used in the computations of 
$F^\ast(\vcc,z)$, for $z\leq u$. 
Thus $P^{*}(|f|, |\vec{x}|,u)= P(|f^{*}_{u}|, |\vec{x}|)$ where 
$f^{*}_{u}(z_{i})=f(z_{i})$ for all $z_{i}$ such that $f(z_{i})$ is used 
in the computations of $F^\ast(\vcc,z)$, $z\leq u$, and $f^{*}_{u}(w)=0$ otherwise.

If $u+1$ is larger than $P^{*}(|f|, |\vec{x}|,u)$ the recursion is 
aborted and the value of $F^\ast(\vcc,w)$ for all $w>u$ is set equal to 
the value $F^\ast(\vcc,u)$. 

\noindent{\bf Claim:}~~Functional 
$\hat{F}(\vcc)$ defined above has the same values as $F(\vcc)$. 

To prove the claim assume $u+1$ is (the first value) larger than $P^{*}(|f|, |\vec{x}|,u)$. 
Then $P^{*}(|f^{*}_{u}|, |\vec{x}|,u)= P(|f^{*}_{u}|, |\vec{x}|)$ and so for 
$f^{*}_{u}$ 
the value of the approximation $P^{*}(|f^{*}_{u}|, |\vec{x}|,u)$ is equal 
to the  value of the second order polynomial $P(|f^{*}_{u}|, |\vec{x}|)$.
However, using the condition 
$$(\forall y) (y \geq P(|f^{*}_{u}|, |\vec{x}|)
\,\impl\, F^\ast (f^{*}_{u},\vec{x}, y)=F^\ast (f^{*}_{u},\vec{x}, P(|f^{*}_{u}|, |\vec{x}|))$$
we get that the recursion procedure produces for all 
$y\geq P(|f^{*}_{u}|, |\vec{x}|)$ the same 
value $F^\ast(f^{*}_{u},\vec{x},P(|f^{*}_{u}|, |\vec{x}|))$. Using the condition
\begin{eqnarray*}
	&(\forall f)(\forall f^*)((\forall y\leq P(|f^{*}|, |\vec{x}|)) (F^\ast(\vcc,y)=
F^\ast(f^*,\vec{x},y))\,\impl &\\
&(\forall y)(F^\ast(\vcc,y)= 
F^\ast(f^*,\vec{x},y)))&
\end{eqnarray*}
we get that indeed 
$$(\forall y)(F^\ast(\vcc,y)= 
F^\ast(f^*_{u},\vec{x},y))$$
This implies that 
$$(\forall y) (y \geq P^{*}(|f^{*}_{u}|, |\vec{x}|) \,\impl\,
F^\ast (\vcc, y)=F^\ast (f^{*}_{u},\vec{x}, P^{*}(|f^{*}_{u}|, |\vec{x}|))$$
Since $ P(|f|, |\vec{x}|)\geq  P^{*}(|f^{*}_{u}|, |\vec{x}|)$ we get 
that 
$$F^\ast (\vcc, P(|f|, |\vec{x}|))=F^\ast (f^{*}_{u},\vec{x}, P^{*}(|f^{*}_{u}|, |\vec{x}|))$$. 
However, the computation of $F^\ast (f^{*}_{u},\vec{x}, P^{*}(|f^{*}_{u}|, 
|\vec{x}|))$ clearly satisfies definition of the computation of 
$\hat{F}(\vcc)$, which proves our claim.

We now formalize in $\BS$ definition of $\hat{F}(\vcc)$. We encode 
the computations for the recursion steps and for approximations 
$P^{*}$ of the second order polynomial $P$ on the basis of values 
$f(z_{i})$ used in the previous steps of recursion.   
Using the fact that 
$$(\forall y \leq P(|f|, |\vec{x}|))(|F^\ast (\vcc, y)|\leq Q(|f|, 
|\vec{x}|))$$ and that 
$(\forall y) (y \geq P(\avc) \,\impl\, F^\ast (\vcc, y)=F^\ast (\vcc, P(\avc))$
we can replace second order polynomial bound $Q(\avc)$ with 
a ${\bf \Sigma}^b_1$ formula with the existential prefix $\exists \vec{z} 
\leq {\vec{t}}^{Q}$ with therms $\vec{t}$ associated with the 
polynomial $Q$, which defines the graph of $\hat{F}(\vcc)$.

By Lemma \paref{PB} it is provable in $\BS$  that there exists $u$ 
such that $u+1$ is larger than $P^{*}(|f|, |\vec{x}|,u)$ 
and so the computation of $\hat{F}(\vcc)$ eventually terminates. 
This implies that $\hat{F}(\vcc)$ is provably total in $\BS$ and thus 
a basic feasible functional. By our claim so is ${F}(\vcc)$.
\qed 

We can now improve the Cook Kapron result on machine models. Recall 
that a functional $\FF$ is {\it computable in polynomial time} 
if there exists an oracle Turing machine $M$ with oracles for 
functions $\vec{f}$ and a second order polynomial $\PP$ 
such that $M$ computes $\FF$ and for all $\vccc$,
the running time $T(\vccc)$, obtained by counting each oracle 
query as a {\it single step} regardless of the size of the oracle 
output, satisfies
$$T(\vccc)\leq \PP.$$ 

\begin{theorem}\label{pbff} {\sl A functional $\FF$ is a polynomial time computable
functional if and only if it is a basic feasible functional.}
\end{theorem} 

\PF By an easy inductive argument on the complexity of the definition of 
functionals one can show that every basic feasible
functional is polynomial time computable (This holds even 
even if we count every oracle
query needed to get $f(z)$ as $|f(z)|$ many steps of computation, 
as it is proved in see \cite{KapCook}). Assume that for a second order 
polynomial $P(\avc)$ a Turing machine halts in $P(\avc)$ many steps. 
Then at each stage of the computation the working tape and the oracle 
input tape can contain at most $P(\avc)$ many symbols, and so the 
oracle output tape can contain at most $|f|(P(\avc))$ many symbols. 
It is easy to see that this implies that the output of such a machine 
can be obtained by polynomially bounded recursion of polynomial 
length, and thus it is a basic feasible functional.

\qed
\section{The Bibliography}

\end{document}